\documentclass[%
 reprint,
 amsmath,amssymb,
 aps,
]{revtex4-1}

\usepackage{graphicx}
\usepackage{dcolumn}
\usepackage{bm}

\begin{document}

\preprint{APS/123-QED}

\title{Maximum Caliber and quantum physics}

\author{Ignacio J. General}
 \altaffiliation{School of Science and Technology, Universidad Nacional de San Martin, and CONICET, 25 de Mayo y Francia, San Martín, 1650 Buenos Aires, Argentina}




\date{\today}

\begin{abstract}
MaxCal is a variational principle that can be used to infer distributions of paths in the phase space of dynamical systems. It has been successfully applied to different areas of classical physics, in particular statistical mechanics in and out of equilibrium. In this work, guided by the analogy of the formalism of MaxCal with that of the path integral formulation of quantum mechanics, we explore the extension of its applications to the realm of quantum physics, and show how the Lagrangians of both relativistic and non-relativistic quantum fields can be built from MaxCal, with a suitable set of constraints. Related, the details of the constraints allow us to find a new interpretation of the concept of inertia. 

\begin{description}
\item[PACS numbers]
01.40.Fk
\end{description}
\end{abstract}

\pacs{Valid PACS appear here}
\maketitle


\section{Introduction}

The \textit{Principle of Maximum Caliber} (MaxCal), first proposed by Jaynes \cite{jaynes-80}, as a generalization of the principle of \textit{Maximum Entropy} (MaxEnt)\cite{jaynes-57-1,jaynes-57-2} has become more popular only in the last two decades\cite{dill-13}, and is starting to prove itself as a powerful tool in physics. It has been shown that MaxCal contains the more usual descriptions of the dynamics of physical systems: the principle of least action, Euler-Lagrange equations and Newton's second law. 

In the next section we present the derivation of the least action principle from MaxCal, and show that the result of MaxCal is more general that just that principle; in fact, MaxCal allows more trajectories than just the one that extremizes the action. This resembles the case of quantum physics, where many trajectories are allowed. But what are those extra trajectories permitted by MaxCal? Making a connection with the path integral formulation of quantum mechanics, we show that the probability of those trajectories is extremely suppressed for classical values of the action, resulting in an agreement of MaxCal and classical physics. 

In the following section, and based on the found relation between MaxCal and quantum mechanics, we explore the relevance of the principle to quantum physics. In particular, we show that both relativistic (Klein-Gordon and Dirac) and non-relativistic (Schr\"odinger) versions of quantum mechanics can be derived from MaxCal with a suitable choice of constraints.

\section{The principle of Least Action as a consequence of MaxCal, and its relation to quantum mechanics}

Maximization of the caliber is a variational  principle  that allows the inference of probability distributions compatible with a set of given constraints \cite{jaynes-80,dill-13}. This principle is a straight generalization of the MaxEnt principle \cite{jaynes-57-1,jaynes-57-2}, where microstates--microscopic realizations of the conformation of a system--are replaced by \textit{microtrajectories} between two points in the phase space--microscopic realizations of the passage of a system from one point to another. 

It has already been shown to be a successful tool to derive several relations related not only to equilibrium but also to systems out of equilibrium. Fick's first and second laws of diffusion, Fourier's law of heat transfer, Newton's second law, Brownian motion, Onsager's  reciprocal relationships, Prigogine’s Principle of Minimum Entropy Production, are some examples of this \cite{wang-05,wang-06,gonzalez-14,dill-18}.

Here, following Wang \cite{wang-06} (with a slightly different point of view), we want to show that the principle of Least Action can be obtained starting from MaxCal, by choosing suitable constraints. Let's take a system that moves from points $a$ to $b$ in its phase space, subjected to the following constraints: 
\begin{enumerate}
  \item Each individual path $i$ between $a$ and $b$ is characterized by a well defined physical property $A_{ab}(i)$, and its average over all possible trajectories is $A_{ab}$. 
  \item The sum of the probabilities for all paths is 1.
\end{enumerate}

Mathematically, we have to maximize the caliber
\begin{equation} 
  S(a,b) = -\sum_{i=1}^{N} p_i (ab) \ln p_i (ab),
\end{equation}
constrained by 
\begin{equation} 
  A_{ab} = \sum_{i=1}^{N} p_i (ab) A_{ab} (i) \text{ \;\;\;\;\;\; and \;\;\;\;\;\; }
  \sum_{i=1}^{N} p_i(ab) = 1. \label{constraints}
\end{equation}
Here, $p_i(ab)$ represents the probability that the system will follow path $i$ when going from points $a$ to $b$ in the phase-space; and $N$ is the number of possible paths connecting those two points. Using the Lagrange multipliers method, we define the auxiliary function
\begin{widetext}
	\begin{equation}
		S' = -\sum_{i=1}^{N} p_i (ab) \ln p_i (ab) - \lambda \left( \sum_{i=1}^{N} p_i(ab) - 1 \right) - \eta \left( \sum_{i=1}^{N} p_i (ab) A_{ab} (i) - A_{ab} \right),
	\end{equation}
\end{widetext}
and calculate stationary points with respect to $p_i$ and the two multipliers, $\lambda$ and $\eta$:
\begin{eqnarray}
  \frac{\partial S'}{\partial p_i} = 0 \Rightarrow & & -\ln p_i -1 -\lambda -\eta A_{ab} (i) = 0 \label{S'} \\ 
  \frac{\partial S'}{\partial \lambda} = 0 \Rightarrow & & \sum_{i=1}^{N} p_i = 1 \label{lambda} \\ 
  \frac{\partial S'}{\partial \eta} = 0 \Rightarrow & & \sum_{i=1}^{N} p_i A_{ab}(i) = A_{ab} \label{eta}
\end{eqnarray}
Combining  Eqs. (\ref{S'}) and (\ref{lambda}), it is possible to eliminate $\lambda$, and arrive at an expression for the probability of path \textit{i}:
\begin{equation}
  p_i (ab) = \frac{1}{Z} e^{-\eta A_{ab} (i)}, \label{P}
\end{equation}
where we define the partition function as 
\begin{equation}
  Z = \sum_{i=1}^{N} e^{-\eta A_{ab} (i)}. \label{Z}
\end{equation}
Notice that 
\begin{equation}
  -\frac{\partial \left( \ln Z \right) }{\partial \eta} = A_{ab},
\end{equation}

Eq. (\ref{P}) shows the relation between the probability of path \textit{i} and its corresponding property $A_{ab}(i)$: path $i$ between $a$ and $b$ has maximum probability, $p_i(ab)$, when $A_{ab}(i)$ is minimum. Notice that if in the last sentence we replace "$A_{ab}(i)$" by the word "action", we are then stating the principle of Least Action, with a caveat: MaxCal (plus constraints in Eqs. (\ref{constraints})) contains the Least Action principle as its most probable outcome, but it is not equivalent to it, since MaxCal allows other trajectories with lower probability. But, since classically there is only one allowed trajectory, we can infer that $\eta$ must be high, thus suppressing the extra trajectories, and leaving only the most probable one. 

To be more specific, consider a simple example, where a system can go from $a$ to $b$ via three possible paths, with values of the action of 1, 2 and 3 (in arbitrary units), respectively. The probability of these cases, calculated via Eqs. (\ref{P}) and (\ref{Z}), are shown in table \ref{action-vs-eta}. There we can see that the probability of the paths with higher action decrease extremely fast if $\eta$ is large. Hence, a classical system could be perfectly described by MaxCal with a large enough value of the multiplier, $\eta$.  

\begin{table}[ht]
	\caption{Probability as a function of the action of three different paths (with values of their action of 1, 2 and 3, in arbitrary units), for different $\eta$ values}
	\label{action-vs-eta}
	\begin{ruledtabular}
    	\begin{tabular}{cccc} 
			$A_{ab}(i) $ & 1 & 2 & 3 \\ 
			\colrule
			$p(i) \, (\textrm{with }\eta=\;\;\;\;1)$ & 0.66 & 0.24 & 0.09 \\ 
			$p(i) \, (\textrm{with }\eta=\;\;10)$ & 0.99 & $10^{-5}$ & $10^{-9}$ \\ 
			$p(i) \, (\textrm{with }\eta=100)$ & 1 & $10^{-44}$ & $10^{-87}$ \\ 
		\end{tabular} 
    \end{ruledtabular}
\end{table}

As noted by Davis and Gonzalez\cite{gonzalez-15}, there is a strong resemblance between the MaxCal formalism and the path integral formulation of quantum mechanics\cite{feynman-hibbs}. More specifically, from the path integral perspective a particle can go from $a$ to $b$ via any physical path connecting them, and each path \textit{i} contributes an exponential factor of the action to the probability amplitude, $P\!A$, of the process:
\begin{equation}
  P\!A_i (ab) \propto e^{\frac{i}{\hbar} A_{ab} (i)}. \label{PATH}
\end{equation}

A comparison with Eq. (\ref{P}) suggests that $\eta$ is inversely proportional to Planck's constant, making its value very large. Hence, when the action is classical (large values of $A$), this would make the probability of non-minimal action trajectories vanish, hence the results of MaxCal agreeing with those of classical physics. 

We have to mention that, although we have compared Eqs. (\ref{P}) and (\ref{PATH}), their meanings are not the same. Eq. (\ref{P}) is the probability of the particle taking path $i$, and, thus, the path with minimum action will be the most likely. Also, as we already said, if $\eta$ is large, other paths will be highly suppressed. 

On the other hand, Eq. (\ref{PATH}) is not the probability but the probability amplitude of path $i$. That is, in quantum mechanics we have to add the $P\!A$s of every path, then take the absolute value and square it in order to get the probability of the particle going from $a$ to $b$ (notice the identity of the path is lost here, as there is no such thing as a precise path in quantum mechanics): 
\begin{equation}
  p(ab) \propto \left| \sum_{i=1}^{N} P\!A_i(ab) \right|^2. \label{PATH2}
\end{equation}
In this case, the largest contribution to $p(ab)$ comes from the path which makes $A_{ab}(i)$ stationary: in the limit of large $A_{ab}(i)/h$ the imaginary exponential oscillates so fast that most paths cancel each other out in the final amplitude; the only path not canceled is the one which makes the action stationary, i.e., the classical path (see \cite{feynman-hibbs} for more details).

\section{Quantum mechanical equations from MaxCal}
Gonzalez \textit{et al}\cite{gonzalez-14} have shown that Newton's second law can be derived from MaxCal imposing two constraints, one related to the magnitude of the squared displacement, and the other related to the probability distribution of the position. Inspired by this, and noticing the above mentioned relation between MaxCal and the path integral formulation of quantum mechanics, we ask ourselves what kind of constraints are needed in order for the different quantum mechanical Lagrangians to appear in the probability of path $i$ as given by MaxCal?

The Lagrangians we propose to find are the non-relativistic Schr\"odinger, and the relativistic Klein-Gordon and Dirac. We start with the latter two, as they are more straight-forward.

\subsection{Klein-Gordon Lagrangian}

\begin{figure}[t]
	\includegraphics[width=3in]{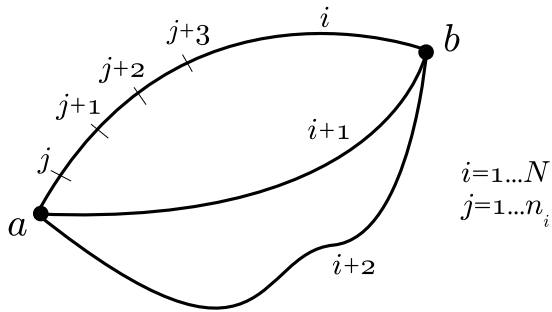} 
	\caption{Points $a$ and $b$ can be connected by $N$ different paths, each designated by an index $i$. And each path's length is discretized in $n_i$ intervals, designated by index $j$}.
	\label{paths_ab}
\end{figure}

Figure \ref{paths_ab} depicts the paths connecting phase space points a and b, where each path $i$ is discretized in time steps $\Delta t$, resulting in $N$ total paths, with $n_i$ number of segments each. 
Let's propose the following constraint, where we use Einstein's notation (repeated indexes in the same term, one as a subscript, the other as a superscript, mean summation with respect to them):
\begin{eqnarray}
  \left\langle \sum _{j=1} ^{n_i} \frac{\Delta \phi_j}{\Delta x_\mu} \frac{\Delta \phi_j}{\Delta x^\mu} \Delta^4 x \right\rangle _i &=& a \label{R1-Klein} \\
  \left\langle \sum _{j=1} ^{n_i} |\phi_j|^2 \Delta^4 x \right\rangle _i &=& b \label{R2-Klein}
\end{eqnarray}

Now, using the Lagrange multipliers technique, we extremize the Caliber, subjected to the above constraints, and impose also that the sum of probabilities must equal 1. That is, we define the function
\begin{widetext}
	\begin{equation}
		S' = \sum _{i=1} ^{N} p_i \texttt{ln}(p_i) - \lambda \sum _{i=1} ^{N} p_i - \beta \sum _{i=1} ^{N} \left( \sum _{j=1} ^{n_i} \frac{\Delta \phi_j}{\Delta x_\mu} \frac{\Delta \phi_j}{\Delta x^\mu} \Delta^4 x \right) - \gamma \sum _{i=1} ^{N} \left( \sum _{j=1} ^{n_i} |\phi_j|^2 \Delta^4 x \right) ,
	\end{equation}
\end{widetext}
and requiring its derivatives with respect to $p_i$ and to the multipliers, $\lambda$, $\beta$ and $\gamma$, to be zero, we obtain the probability of path $i$:
\begin{equation}
	p_i(ab) = \frac{1}{Z} e^{-A_{ab}(i)}, \label{pi}
\end{equation}
where 
\begin{equation}
	A_{ab}(i) = -\beta \sum _{j=1} ^{n_i} \left( \partial _\mu \phi_j \right) \left( \partial ^\mu \phi_j \right) \Delta^4 x - \gamma \sum _{j=1} ^{n_i} |\phi_j|^2 \Delta^4 x, \label{S-Klein}
\end{equation}
and
\begin{equation}
	Z = \sum_{i=1}^{N} e^{-A_{ab}(i)},
\end{equation}
Notice in Eq. (\ref{S-Klein}) we have replaced the rates of change of the field ${\Delta \phi_j}/{\Delta x^\mu}$ by the derivatives $\partial_\mu \phi_j$. $Z$ is a normalization factor.

Finally, recalling Eq. (\ref{P}) and comparing with Eq. (\ref{pi}), we see that $A$ is the action. And since the action is related to the Lagrangian through $A = \int L \, d^4x$, we can conclude--after transforming from the discrete to continuous case--that the corresponding Lagrangian is
\begin{equation}
	L = \frac{1}{2} \left(\partial_\mu \phi \right) \left(\partial^\mu \phi \right) - \frac{1}{2} m^2 |\phi|^2. \label{L-Klein}
\end{equation}
In the last step we assigned the following values to the multipliers, in order to recover the well-known Klein-Gordon Lagrangian: 
\begin{eqnarray}
	\beta &\rightarrow& -1/2 \\
    \gamma &\rightarrow& m^2/2
\end{eqnarray}
We see that the mass is associated with the constraint in Eq. (\ref{R2-Klein}), via the $\gamma$ multiplier (it has to be this way, since in relativistic field theories, the Lagrangian term quadratic in the field is the one that gives mass to the field). To interpret this, let's recall from the theory of Lagrange multipliers \cite{karabulut-06} that the multiplier has a specific meaning: it is the derivative of the function being extremized (the caliber $S$) with respect to the value of the constraint [$b$ in Eq. (\ref{R2-Klein})]: $m^2 \sim \partial S / \partial b$. Hence, the mass of the field is associated with the rate of change of the caliber with respect to $b$. That is, a large $m$ signifies that changing $b$ would largely change $S$, getting it out of the stationary point (in which it \textit{wants} to stay, according to MaxCal). On the contrary, a field with a small $m$ would barely change the value of $S$ when varying $b$. In conclusion, the mass has to do with the slope of the curve $S$ vs $b$: zero slope associated with zero mass, and increasing slopes associated with increasing mass. Interpreting this in a more intuitive physical way, a field with zero mass could be thought of as having no constraints on the values the field can take in different points of space-time. A massive field, on the other hand, cannot take any value. This suggests an interpretation of the concept of inertia: The value of the field of a large mass can not change easily, since the change in $S$ would be large. Hence, the field (or particle) can only change its state of motion slowly, with a large inertia. But a light particle can move fast and change its path easily, since it has no cost to change the values of the field (thus, small inertia).

\subsection{Dirac Lagrangian}
Analogously, we now propose constraints which will lead to the Dirac Lagrangian ($L = i \bar{\Psi} \gamma^\mu \partial_\mu \psi - m \bar{\psi} \psi$). These are:
\begin{eqnarray}
  \left\langle \sum _{j=1} ^{n_i} \bar{\psi}_j \gamma^\mu \frac{\Delta \psi_j}{\Delta x_\mu} \Delta^4 x \right\rangle _i &=& a \label{R1-Dirac} \\
  \left\langle \sum _{j=1} ^{n_i} \bar{\psi}_j \psi_j \Delta^4 x \right\rangle _i &=& b, \label{R2-Dirac}
\end{eqnarray}
along with the constraint of total probability equal to 1. Applying Lagrange multipliers to extremize the caliber, we arrive at an expression for the probability of path $i$ as that in Eq. (\ref{pi}), but with the action now given by
\begin{equation}
	A_{ab}(i) = -\beta \sum _{j=1} ^{n_i} \bar{\psi} \gamma^\mu \left( \partial_\mu \psi \right) \Delta^4 x - \gamma \sum _{j=1} ^{n_i} \bar{\psi} \psi \Delta^4 x. \label{S-Dirac}
\end{equation}
Or, going to the continuous limit,
\begin{equation}
	L = i \bar{\psi} \gamma^\mu \left( \partial_\mu \psi \right) - m \bar{\psi} \psi, \label{L-Dirac}
\end{equation}
where we have assigned the following values to the multipliers: 
\begin{eqnarray}
	\beta &\rightarrow& -i \\
    \gamma &\rightarrow& m.
\end{eqnarray}
We see that in this case it is $m$ that is associated with the cost of having defined values of the square of the field, and not $m^2$ as in the previous case. This difference seems interesting, but shouldn't be surprising, since the fields themselves are intrinsically different, with Klein-Gordon fields being scalar and Dirac fields vectorial. 

\subsection{Schr\"odinger Lagrangian}
We could repeat the above process again, this time to show that the Schr\"odinger Lagrangian can also be obtained by proposing suitable constraints. These are
\begin{eqnarray}
  \left\langle \sum _{j=1} ^{n_i} \frac{\Delta \psi}{\Delta x_j} \frac{\Delta \psi^*}{\Delta x_j} \Delta t \right\rangle _i &=& a \label{R1-Schro} \\
  \left\langle \sum _{j=1} ^{n_i} \psi^* \frac{\Delta \psi}{\Delta t}  \Delta t \right\rangle _i &=& b \label{R2-Schro} \\
  \left\langle \sum _{j=1} ^{n_i} \psi^* \psi \Delta t \right\rangle_i &=& c. \label{R3-Schro} 
\end{eqnarray}
And again, using Lagrange multipliers and taking the continuous limit, we get
\begin{equation}
	L = \beta \left(\partial_j \psi \right) \left(\partial_j \psi^* \right) + \gamma \psi^* \dot{\psi} + \delta \psi^* \psi. \label{L-Schro}
\end{equation}
The assignment of multipliers this time is:
\begin{eqnarray}
	\beta &\rightarrow& \frac{h^2}{8\pi^2 m} \\
    \gamma &\rightarrow& \frac{h}{4\pi i} \\
    \delta &\rightarrow& V.
\end{eqnarray}
Not surprisingly, mass is not associated in this case with the quadratic term in the field, but with the one with the spatial derivatives. In general, relativistic theories like Klein-Gordon and Dirac have mass provided by the Lagrangian term that is quadratic in the field; non-relativistic theories, like Schr\"odinger, have mass in the terms with spatial derivatives. This is a consequence of relativistic and non-relativistic theories obeying different dispersion relations: $E^2=p^2+m^2$ and $E=p^2/2m$, respectively. Recalling that in quantum mechanics the $\hat{p}$ operator is associated with spatial derivatives, we see the reason why the Sch\"odinger Lagrangian has $m$ dividing them.

\section{Conclusion}
In this work we have shown, first, that the principle of least action follows from MaxCal, with the constraint that the value of the action in each path and the average action are both known. Notice that we haven't shown why the constraint has to be on the action, but we have taken the action as a fundamental quantity that describes (constrains) the different paths. But MaxCal is more general than Least Action, as it allows not only the classical path which extremizes the action and is thus the most likely, but gives non-zero probability for other paths [Eq. (\ref{P})]. This fact leads to a connection between MaxCal and the path integral formulation of quantum mechanics, as probability amplitudes in that formalism are also described by exponential functions of the action [Eq. (\ref{PATH})]. Unless further work says otherwise, this should be taken more as an analogy than as a direct connection, since the relation we mentioned is between \textit{probability} of a specific path $i$, in MaxCal, and \textit{probability amplitude} of a specific path $i$, in the path integral formulation. Those concepts, though firmly related [through Eq. (\ref{PATH2})], are not the same.

Inspired by this similarity, we probed the relevance of MaxCal to quantum mechanics, in particular, the possibility of obtaining quantum equations from that principle. And, in fact, we showed that the Klein-Gordon, Dirac and Schr\"odinger Lagrangians can be obtained from MaxCal, by imposing constraints on the average value of products of the field and/or its derivatives [see constraint Eqs. (\ref{R1-Klein}), (\ref{R2-Klein}), (\ref{R1-Dirac}), (\ref{R2-Dirac}), (\ref{R1-Schro}), (\ref{R2-Schro}) and (\ref{R3-Schro})]. The procedure we followed in order to get them was backwards, i.e., knowing the Lagrangians, we proposed constraints that would lead to them. 

An interesting finding that arises from the constraints is that related to the mass of the fields. In the relativistic cases, we have seen that the mass of the field is generated from the constraints in Eqs. (\ref{R2-Klein}) and (\ref{R2-Dirac}): from the theory of Lagrange multipliers, mass is the cost of having the average of the squared fields set to a specific value. A large mass means that the specific value of $b$ in the constraint can hardly be changed, while a massless field would signify the value of $b$ is not important, and can easily be changed. In turn, this appears as an explanation of the concept of inertia, that is, it is difficult to change the values of the field (and thus, the path of the particle) when it has a large mass, but not when it is light. 

Related to the other constraints used in the quantum cases, one may wonder what is special about those constraints? Why not others? To answer this question we rely on the fact, well-known in field theory, that Lagrangians need to be covariant and, thus, their terms need to be scalars. In this way, the terms available to a constraint (from which the Lagrangian will be generated) are only those that are scalar, for example, combinations of field and derivatives where all indices are summed over. If we also require simple combinations of field and first derivatives (higher order derivatives are usually not needed in most theories), then the possible terms are highly restricted to those used in our constraints. 

In summary, this work shows that the MaxCal principle, when complemented with the right set of constraints, is useful not only in classical mechanics, but also in quantum theory, as it allows the derivation of relevant Lagrangians, like Schr\"odinger, Klein-Gordon and Dirac's. Moreover, exploring the nature of the constraints and of the associated Lagrange multipliers may lead, as we have shown in relation to the mass, to new insights into important physical concepts. 

Finally, an intriguing question arises from the similarity of Eqs. (\ref{P}) and (\ref{PATH}). Is their relation just qualitative, as we have used it here? It would be interesting to see if there is a way to firmly connect one to the other, thus logically connecting classical to quantum mechanics.

\section*{Acknowledgements} 
The author acknowledges support from Agencia Nacional de Promoci\'on Cient\'ifica y Tecnol\'ogica, for grant PICT-2015-3832.


%

\end{document}